\documentclass[letterpaper,twoside,english]{article}
\usepackage[T1]{fontenc}
\usepackage[latin1]{inputenc}
\usepackage{graphicx}
\usepackage[authoryear]{natbib}

\makeatletter

\providecommand{\tabularnewline}{\\}

\usepackage{graphicx}
\usepackage{latexsym,times}
\usepackage[intlimits]{amsmath}
\usepackage{amsfonts,amssymb}
\DeclareSymbolFontAlphabet{\mathbb}{AMSb}
\usepackage[mathscr]{eucal}
\usepackage{lineno}
\usepackage[textstyle,cdot,squaren,thickspace,thickqspace]{SIunits}
\usepackage{fancyvrb}
\usepackage{url}
\usepackage{xspace}
\usepackage{psfrag}
\usepackage[ps2pdf]{thumbpdf}              
\usepackage[%
bookmarks=true,
bookmarksnumbered=true,
hypertexnames=false,
pdfborder={0 0 0},
ps2pdf]{hyperref}%
\hypersetup{
  pdfauthor   = {can ozan tan <tanc@cns.bu.edu>},
  pdftitle    = {},
  pdfsubject  = {},
  pdfkeywords = {},
  pdfcreator  = {LaTeX with hyperref package},
  pdfproducer = {dvips and ps2pdf}
}
\usepackage[letterpaper,top=1in,bottom=1in,hmargin={1in,1in}]{geometry}
\numberwithin{equation}{section}
\usepackage{float}
\usepackage[bf,float,subfigure]{caption2}
\setcaptionmargin{0.5in}
\usepackage[sf]{subfigure}

\DeclareRobustCommand\em
{\@nomath\em \ifdim \fontdimen\@ne\font >\z@
  \upshape \else \slshape \fi}

\usepackage{babel}

\usepackage{babel}
\makeatother
\begin{document}
\date{}

\title{\normalsize \textit{Running Head:} Predictive Models in Ecology \\ \Large Predictive Models for Characterization of Ecological Data}

\author{Can Ozan Tan$^{1*}$, Uygar \"{O}zesmi$^{2}$ and Bahtiyar Kurt$^{3}$}

\maketitle
\flushleft

\noindent $^{1}$Canakkale Onsekiz Mart University, Department of
Biological Sciences, 17020, Canakkale, Turkey

\noindent $^{2}$Erciyes University, Dept. of Environmental Engineering,
38039 Kayseri, Turkey

\noindent $^{3}$Doga Dernegi P.O. 640, Yenisehir, Ankara TR-06445,
Turkey

\noindent $^{*}$Corresponding author: Boston University, Department
of Cognitive and Neural Sciences, 677 Beacon St. \#201, Boston, MA
02215. Tel: +1 617 353-6741, Fax: +1 617 353 7755 e-mail: tanc@cns.bu.edu

\begin{abstract}

Although ARTMAP and ART-based models were introduced in early 70's they were not used in characterizing and classifying ecological observations. ART-based models have been extensively used for classification models based on satellite imagery. This report, to our knowledge, is the first application
of ART-based methods and specifically ARTMAP for predicting habitat
selection and spatial distribution of species. We compare the performance of ARTMAP to assess the breeding success of three bird species (\textit{Lanius senator}, \textit{Hippolais pallida}, and \textit{Calandrella brachydactyla}) based on multi-spectral satellite imagery and environmental variables. ARTMAP is superior both in terms of performance (percent correctly classified - $pcc = 1.00$) and generalizability ($pcc >0.96$) to those of feedforward multilayer backpropogation ($>0.87$, $>0.65$), linear and quadratic discriminant
analysis ($>0.48$, $>0.46$) and k-nearest neighbor ($>0.82$, $>0.66$) methods. Compared to other methods, ARTMAP is able to incorporate new observations with far less computational effort and can easily add data to already trained models.

\noindent \textbf{Keywords:} ART; ARTMAP; artificial neural networks; backpropogation; pattern recognition; spatial habitat selection; \textit{Lanius senator;} \textit{Hippolais pallida;} \textit{Calandrella brachydactyla.}
\end{abstract}
\newpage
\linenumbers

\section{Introduction}

Characterization of observations to explain interactions in an ecosystems
as well as within communities and individual species, in order to predict
a state has been one of the main problems in ecology. The inherent
complexity of the ecological processes, the relatively limited number
of possible observations and their susceptibility to observational
and/or measurement noise has been considered among the major difficulties
in predicting a state in ecology \citep{fielding99}. Subject to these
constraints, efforts to characterize ecological data and predict the
state of a given ecosystem or community shifted towards statistical
methods, rather than box-and-arrow type differential equation models
\citep{ross1976,lassiter:kearns1977}. Statistical models proved to
be more robust in terms of capturing nonlinearities and being generalizable
over new data sets \citep{moilanen1999,devalpine2003}.

Several statistical techniques are readily available for the use of
ecologists to characterize observations. These range from simple regression
models \citep{gutierrez:etal2005,miller2005} to generalized additive
\citep{dunk:etal2004} and linear models \citep{ozesmi:mitsch1997,tan:beklioglu2005}
and from classification algorithms such as k-nearest neighbor (k-NN),
linear and quadratic discriminant analysis (LDA and QDA, respectively)
\citep{joy:death2003,maron:mill2004} to recently genetic algorithms
\citep{underwood:etal2004}, pattern recognition methods, such as
artificial neural networks \citep{lek:etal96,recknagel:etal97,lek:guegan99,ozesmis99}
and lately ecological data mining \citep{chawla:etal2001}. While
standard parametric methods such as LDA, QDA and regression are mostly
criticized as being dependent on strong assumptions about the distribution
of the underlying data \citep{hastie:etal01}, classification and
pattern recognition methods require large number of training points.
On the other hand, artificial neural network-based approaches are blamed
to be black-box models thus not being able to provide insight into
the complex interactions of the ecosystem processes, although they
are able to overcome the difficulties associated with traditional
statistical models \citep{bishop95,ripley96,hastie:etal01}. Nevertheless,
artificial neural network-based models can provide valuable insight
into ecosystem dynamics as there are several techniques for 'opening
the black-box' \citep{ozesmis99,olden:jackson2000,ozesmi:etalinpress}.

Recently, backpropogation based methods became popular in ecological
applications. Their use range from characterization of habitat selection
of phytoplankton \citep{scardi96,scardi01} to fish \citep{reyjol:etal01}
and bird species \citep{ozesmis99}, to modeling whole communities
and ecosystems \citep{tan:smeins96,tan:beklioglu2005} and characterization
of wildlife damage \citep{spitz:lek1999} to gain insight into the
dynamical structure of the ecosystems. However, the main drawback
of backpropogation based methods has been that they are inherently
off-line, that is iterative, methods using all the available data
at once. In other words, each time a new observation is made, these
models require to be retrained with the whole data set in order to
include the new observation, thus requiring a significant amount of
computational resources and time. In addition, the fact that the performance,
particularly generalizability, of these methods reduces significantly
with limited number of data points renders this approach to be impractical,
at least in ecology where the number of observations are commonly
limited.

This report aims to introduce another statistical pattern recognition
model, ARTMAP, based on adaptive resonance theory (ART) \citep{grossberg1976a,grossberg1976b},
which is relatively unfamiliar to the ecological community. ART is
originally developed to explain cortico-cortical interactions for
object recognition and learning in the brain during early 70's \citep{grossberg1976a}.
During 80's and early 90's, ART was extended as a pattern recognition
and classification algorithm, and successfully applied to several
benchmark technological data sets and classification of satellite
imagery data \citep{grossberg1988,carpenter:etal1991c,carpenter:etal1997}.
However, despite its long history as a statistical pattern recognition
and classification algorithm, this report, to our knowledge, is the
first application of an ART based algorithm to an ecological data
set. In addition to being on-line (that is a non-iterative learning 
algorithm, which enables easy and fast incorporation of new observations
to an already trained model), ARTMAP also performs significantly better
on the data set considered here, utilizing a considerably smaller
amount of computational time. To that end, we used satellite-based
multi-spectral data and environmental variables to predict the occurrence
of three bird species of Southeastern Anatolia, namely woodchat shrike
\textit{Lanius senator} (Linnaeus, 1758), olivaceous warbler \textit{Hippolais
pallida} (Ehrenberg, 1833), and short-toed lark \textit{Calandrella
brachydactyla} (Leisler, 1814). To predict the occurrence of the three
bird species we used k-NN, LDA, QDA, feedforward multilayer backpropogation
network, and ARTMAP. We provide a discussion of comparative performances
of these different models.

\section{Methods}

\subsection{Traditional Classification Methods}

We compared the performance of fuzzy ARTMAP model against traditional
classification and pattern recognition methods commonly employed in
ecological studies. The first method was k-nearest neighbor method,
which is an accepted benchmark classification method, if one considers
only the training data. Nearest neighbor methods use those observations
in the training set $\mathcal{{T}}$ closest in the input space to
$x$ to form $\hat{Y}$. More specifically,

\begin{equation}
\hat{Y}=\frac{1}{k}\sum_{x_{i}\in N_{k}(x)}y_{i}\end{equation}

\noindent where $N_{k}(x)$ is the neighborhood of $x$ defined by
the $k$ closest points $x_{i}$ in the training sample. It is clear
that when the neighborhood $k$ is considered to be $k=1$, k-NN methods
potentially can reach the minimum classification error possible on
the training set. Note that in this case the error on independent
test set is intuitively expected to be quite high. In addition, we
also used LDA and QDA, which are mostly argued to be \char`\"{}amazingly
robust\char`\"{} on industrial data sets \citep{hastie:etal01}. LDA
and QDA techniques enable one to infer the posterior probabilities
of the output categories based on the data observed, using Bayes theorem:

\begin{equation}
\mathrm{P}(G=k|X=x)=\frac{f_{k}(x)\pi_{k}}{\sum_{l=1}^{K}f_{l}(x)\pi_{l}}\end{equation}

\noindent where $f_{k}(x)$ is the class-conditional density of $X$
in class $G=k$, and $\pi_{k}$ is the prior probability of class
$k$ with $\sum_{k=1}^{K}\pi_{k}=1$. LDA and QDA assume Gaussian
distribution for class densities. Fundamentally, for two category
cases (as in our case), and assuming that the covariances $\Sigma_{k}$
of the class densities are equal, linear discriminant function is
given as

\begin{equation}
\delta_{K}=x^{T}\Sigma^{-1}\mu_{k}-\frac{1}{2}\mu_{k}^{T}\Sigma^{-1}\mu_{k}+\log\pi_{k}\end{equation}

\noindent where the parameters of the Gaussian distributions are estimated
from the data as

\begin{eqnarray}
\hat{\pi}_{k} & = & \frac{N_{k}}{N}\\
\hat{\mu}_{k} & = & \frac{\sum_{g_{i}=k}x_{i}}{N_{k}}\\
\hat{\Sigma} & = & \frac{\sum_{k=1}^{K}\sum_{g_{i}=k}(x_{i}-\hat{\mu}_{k})(x_{i}-\hat{\mu}_{k})^{T}}{(N-K)}\end{eqnarray}

\noindent where $N_{k}$ is the number of class-$k$ observations.
An equivalent decision rule is given as $G(x)=\mathrm{arg}\max_{k}\delta_{k}(x)$.
If the equality assumption of class covariances $\Sigma_{k}$ does
not hold, we obtain quadratic discriminant function

\begin{equation}
\delta_{k}(x)=-\frac{1}{2}\log|\Sigma_{k}|-\frac{1}{2}(x-\mu_{k})^{T}\Sigma_{k}^{-1}(x-\mu_{k})+\log\pi_{k}\end{equation}

\noindent with an equivalent decision boundary between each pairs
of classes $k$ and $l$ described by a quadratic equation $\{ x:\delta_{k}(x)=\delta_{l}(x)\}$.
A more in-depth discussion of these two methods, among with k-NN method,
can be found in \citet{hastie:etal01}.

Traditional classification methods has been often criticized as they
require strong assumptions about the underlying distribution of the
observations \citep{ripley96,hastie:etal01}. To overcome this problem,
connectionist artificial neural network based approaches, such as
feedforward multilayer backpropogation network has become recently
popular among ecological modeling \citep{scardi96,scardi01,tan:beklioglu2005}.
Although ART and ARTMAP family of models are another type of artificial
neural networks, they differ from connectionist approaches in several
aspects \citep{carpenter:etal1991a,carpenter:etal1991b,carpenter:etal1991c,carpenter:etal1992}.
For that reason, we also compared the performance of fuzzy ARTMAP
model to that of a generalized linear model (GLM) and of a multilayer
feedforward backpropogation model.

\subsection{ARTMAP}

\begin{figure}
\begin{center}\includegraphics[%
  width=5in,
  keepaspectratio]{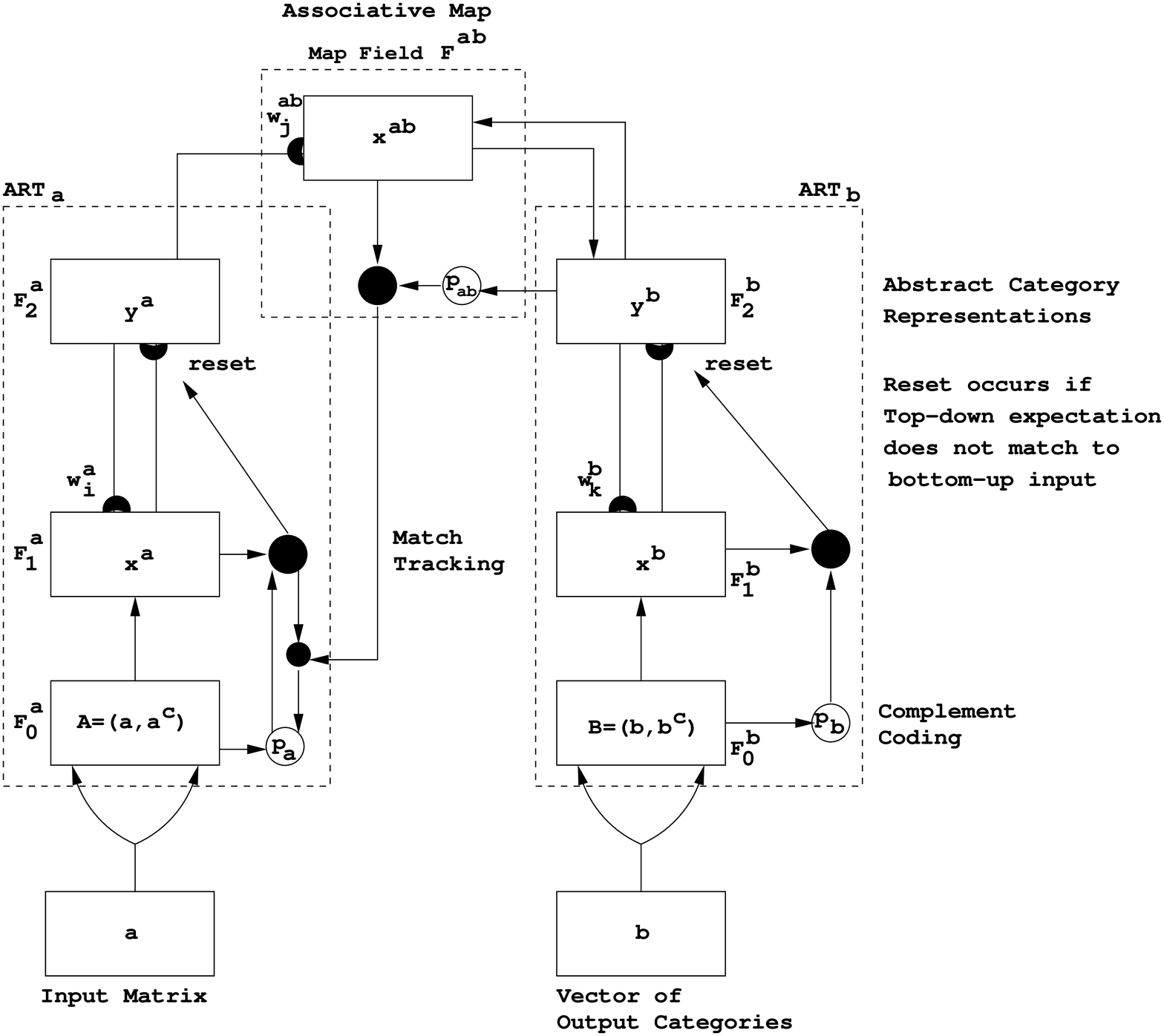}\caption{}\label{fig:fart}\end{center}
\end{figure}

Briefly, ARTMAP architecture consists of two ART modules, which are
self-organizing maps \citep{carpenter:etal1991a}, one for input space
and one for output space (figure \ref{fig:fart}; ART$_{a}$ and ART$_{b}$,
respectively). Learning occurs for each ART module independently,
whenever an expected category matches to presented input pattern,
or a novel input pattern is encountered, then categories are formed
in both ART modules and mapped on an associative learning map field.
Thus, ARTMAP models represent a \char`\"{}pseudo-supervised\char`\"{}
learning method \citep{carpenter:etal1991a}. There are several variants
of ART modules \citep{carpenter:grossberg1990,carpenter:etal1991b,carpenter:etal1991c}.
Here, we used fuzzy ART modules, which were developed as pattern recognition
methods for data sets with continuous input space \citep{carpenter:etal1991c,carpenter:etal1992}.
Shortly, each fuzzy ART system contains an input field $F_{0}$, a
$F_{1}$ field receiving bottom-up signals from $F_{0}$ and top-down
input from $F_{2}$, the latter of which represents the active category
(figure \ref{fig:fart}). So-called complement coding \citep{carpenter:etal1992}
should be employed before feeding the input vectors to fuzzy ART modules.
Theoretical considerations for this requirement are discussed in detail
in \citet{carpenter:etal1992}. Fundamentally by complement coding,
it is meant that an $N\times P$-dimensional input matrix $\mathbf{a}$
is coded and fed to the model as an $N\times2P$-dimensional matrix
$[\mathbf{a},\mathbf{a}^{c}]$, where $a_{i}^{c}=(1-a_{i})$.

At each $F_{2}$ category node, there is a weight associated with
that node, which are initially set to 1. Each weight $w_{ji}$ is
monotonically increasing with time and hence its convergence to a
limit is guaranteed \citep{carpenter:etal1991c,carpenter:etal1992}.
Fuzzy ART dynamics depend on a choice parameter $\alpha>0$, a learning
rate $\beta\in[0,1]$, and a vigilance parameter $\rho\in[0,1]$.
For each given input pattern and $j$th node of $F_{2}$ layer, the
choice function $T_{j}$ is defined by

\begin{equation}
T_{J}{(\mathbf{I})}=\frac{|\mathbf{I}\wedge\mathbf{w}_{j}|}{\alpha+|\mathbf{w}_{j}|}\end{equation}

\noindent where $\wedge$ is the fuzzy AND operator and is equivalent
to component-wise $\min$ operator, $|\cdot|$ is the euclidean norm,
and $\mathbf{w}_{j}=(w_{j1}\cdots w_{jM})$. The system makes a category
choice when at most one $F_{2}$ node can become at a given time,
and the category choice is given as $T_{J}=\max\{ T_{j}:j=1\dots N\}$.
In a choice system, the activity of a given node at $F_{1}$ layer
is given as $\mathbf{x}=\mathbf{I}$ if $F_{2}$ node is inactive
and $\mathbf{x}=\mathbf{I}\wedge\mathbf{w}_{J}$ if $J$th $F_{2}$
node is selected. Resonance occurs in the ART module if

\begin{equation}
\frac{|\mathbf{I}\wedge\mathbf{w}_{J}|}{|\mathbf{I}|}\geq\rho\label{eq:resonance}\end{equation}

\noindent and reset occurs otherwise. If reset occurs, the value of
the choice function $T_{J}$ is set to 0, and a new index $J$ is
chosen. The search process continues until the chosen $J$ satisfies
the resonance criterion (equation \ref{eq:resonance}). Once search
ends and resonance occurs, the weight vector $\mathbf{w}_{J}$ is
updated by

\begin{equation}
\mathbf{w}_{J}^{(\mathrm{new})}=\beta\left(\mathbf{I}\wedge\mathbf{w}_{J}^{(\mathrm{old})}\right)+(1-\beta)\mathbf{w}_{J}^{(\mathrm{old})}\end{equation}

As briefly mentioned above, fuzzy ARTMAP model consists of two fuzzy
ART modules, one for input and one for target vectors linked by an
associative learning network and an internal controller. With reference
to figure \ref{fig:fart}, when a prediction by ART$_{a}$ module,
which receives the input vectors, is disconfirmed at ART$_{b}$ module,
receiving target vector, inhibition of map field activation induces
the match tracking process, which raises the ART$_{a}$ vigilance
$\rho_{a}$ to just above the $F_{1}^{a}$ activation so that the
activation of $F_{0}^{a}$ matches the reset criterion (i.e., $\rho_{a}$
is decreased just to miss the match criterion given by equation \ref{eq:resonance}).
This triggers an ART$_{a}$ search process which leads to activation
of either an ART$_{a}$ category that correctly predicts b at match
field, or to activation of a new node which has not used before (that
is, either an already formed category that predicts $b$ is selected,
or a new category is created). ART and ARTMAP algorithms, in essence,
are similar to k-NN methods with adaptive update of the size of the
neighborhood with each pattern encountered in the data. It is, nevertheless,
a nonlinear algorithm such that the shape of the clusters built based
on the patterns embedded in the input space are nonlinear. For details
of fuzzy ART algorithm as well as for its geometrical interpretation,
readers are referred to \citet{carpenter:etal1991c}, and the details
of fuzzy ARTMAP algorithm can be found in \citet{carpenter:etal1992}.

Although new to ecology, ART and ARTMAP theory has been developed
since early 70's, and the reader is referred to \citet{cohen:grossberg1983}
and \citet{grossberg1988} for theoretical considerations. Generic
implementational issues can be found in \citet{carpenter2003}.

\subsection{Implementation Details}

\subsubsection{Data}

Ornithological and ecological data used in this study has been obtained
from the GAP biodiversity research project of Turkish Society for
the Conservation of Nature (DHKD) conducted between 2001 and 2003
\citep{welch2004}. Detailed description of observations and data
collection method can be found in \citet{bahtez} and \citet{welch2004}.

During the field studies, which lasted two years, 1592 points were
visited and the ecological variables as well as the breeding success
of bird species were recorded. Satellite imagery used in this study
was obtained by the Turkish Society for the Conservation of Nature,
and consisted of LANDSAT images bands 1-5 and 7, with a resolution
of $30\times30$ m. The characteristics of the satellite images and
the properties of the bands used are given in detail in \citet{per2003}
and \citet{bahtez}.

Independent variables were 6 image bands and 6 environmental variables.
Environmental variables were elevation (m), distance to nearest road
(m), distance to water (m), vegetation index (categorical), annual
relative humidity (\%), and annual mean temperature ($^{o}$C). For
all the models considered, the output classes for each data pattern
has been assigned either 0 or 1, depending on the occurrence of individuals
recorded for each bird species considered here \citet{bahtez}.

It is important for statistical learning methods to have an input
space where the number of data points for each output category (0
and 1, in our case) is approximately balanced to avoid biased estimates
\citep{ripley96}. To that end, although there were 1592 data points
collected in our data set, the number of data points corresponding
to category 1 (i.e., the presence of individuals) were limited (246
- 274, depending on the species), and in order to establish balance,
we randomly selected an equal number of data points with output category
0 to the number of points with breeding individuals (category 1) \citep{hirzel:etal2002}.
Thus, the data fed to the models were consisting of 492-548 observations
depending on the bird species considered.

The importance of setting aside independent test data, which should
not be included during training, to assess the actual performance
of a given model has been rigorously emphasized elsewhere \citep{ripley96,ozesmis99,hastie:etal01,tan:beklioglu2005}.
To that end, we randomly split the data sets for each species into
two sets with equal number of data points such that the number of
data points corresponding to each category were still balanced, and
used one set to train the models, while the other to asses the generalizability
of the trained models.

\subsubsection{Traditional Classification Models}

k-NN, LDA, and QDA models were implemented in R-language statistical
software \citep{rlang}. The theoretical considerations and implementation
details for these models can be found in \citet{hastie:etal01}. GLM
and backpropogation models were implemented using NevProp3 software
\citep{goodman96:software}. For backpropogation models, the architecture
of the network is optimized step-wise \citep{ozesmi:etalinpress},
and the networks with 8, 3 and 10 hidden units were used as final
models for \textit{L. senator}, \textit{H. pallida} and \textit{C.
brachydactyla}, respectively. Theoretical considerations for feedforward
multilayer backpropogation networks can be found in \citet{rumelhart86},
\citet{bishop95} and \citet{ripley96}, and the implementation details
of GLM and backpropogation models for this particular study are given
in \citet{bahtez}.

\subsubsection{ARTMAP}

ARTMAP was implemented in Matlab version 7 (Mathworks Inc.). All input
variables were standardized to zero mean, and units of standard deviation
before being fed to all models, but ARTMAP. For ARTMAP, the input
variables are standardized such that they are squeezed into a hypercube
$C^{P}\in[0,1]$, where $P$ is the number of independent features
(i.e., dimension of input space). Theoretical considerations for the
reason to use this particular standardization for ARTMAP models is
beyond the scope of this report, and interested readers are referred
to \citet{kosko1992a}. 

All six models have been trained three times separately for the three
bird species, and each trained model is then tested separately on
corresponding test sets to asses its generalizability. All models
have been trained using bootstrapping and cross-validation to optimize
so-called bias-variance trade-off \citep{hastie:etal01}.

\section{Results and Discussion}

\subsection{Performance of the Models on Training and Independent Tests}

The performances of all five models for all three different bird species
on both training and test sets are given in Table \ref{tab:split}.
For backpropogation models, the performance is given as c-index, which
is approximately the area under the ROC curve \citep{bishop95}. For
other four models, the performance is given as percent correctly classified.
Note that for data sets with perfectly balanced number of data points
corresponding to each output category, percent correct measure is
equivalent to the c-index measure \citep{bishop95,ripley96}. Hence,
the performance measures of all five methods in our case are compatible.
Further note that unlike traditional performance measures such as
$R^{2}$, a value of 0.5 for percent correct and c-index indicates
a performance not better than random.

As evident from Table \ref{tab:split}, the performance of neural
network models, both ARTMAP and backpropagation, was superior compared
to the traditional classification algorithms. For the latter group,
especially for LDA and QDA, the data corresponding to \textit{H. pallida}
seems to be particularly \char`\"{}difficult\char`\"{}, with both
models' performance on training set being around random chance level.
Among traditional classification models, although k-NN performed better
on training set compared to LDA and QDA, it too suffered from low
performance on independent test sets.

Backpropogation and GLM method's performance on training sets was
considerably better than previous three techniques, and it is especially
noteworthy that backpropogation model predicted all of the data points
on the training sets correctly for the data sets of \textit{L. senator}
and \textit{C. brachydactyla}. However with respect to training sets,
ARTMAP model performed same on these two data sets, and better on
set \textit{H. pallida} than backpropogation model. To this end, also
note the number of hidden units in backpropogation and the number
of formed categories at fuzzy ART module for input vectors (committed
nodes) in ARTMAP models (8,3,10 and 2,4,3, respectively). The number
of hidden units (or equivalently, of committed nodes) indicate how
well the input space is represented as a compressed code in the internal
structure of the model \citep{ripley96,carpenter:etal1991a}. Considering
the fact that the number of compressed representations are equivalent
to the degrees of freedom of the model \citep{bishop95,ripley96},
ARTMAP appears to be more effective in representing the input space,
compared to backpropogation method. And it does so without sacrificing
the performance on the training set. In addition, the less the degrees
of freedom of a model is, the more generalizable it would be \citep{hastie:etal01}.
The performances of GLM, backpropogation and ARTMAP models on independent
test sets also revealed this fact in that the predictive power of
ARTMAP was considerably better than the other two, being close to
1 for each of the three independent test cases (Table \ref{tab:split}).
Thus, at least for the current data set considered, ARTMAP seems to
be more robust in characterizing ecological data and predicting the
species occurrence, in terms of both training accuracy and generalizability.

\subsection{Computational Efficiency}

In addition to its superiority in terms of training and test performance,
ARTMAP also has the advantage of being computationally much less expensive
than feedforward backpropogation networks. For the results presented
in this report, backpropogation network required close to 1000 iterations
on the complete training set, which approximately took 18 minutes
on a P4 1.8GHz PC. Noting that backpropogation models also require
architecture as well as free parameter (e.g. learning rate, momentum
etc.) optimization, with each model to be trained separately, to achieve
best performance, the amount of computational time required grows
significantly. On the other hand, fast-learning mode of ARTMAP \citep{carpenter:etal1992}
enables the network to learn \char`\"{}one-shot deals\char`\"{}, that
is to learn without iterating the training set. ARTMAP model on fast-learning
mode on the same system took approx 10 seconds to train and achieve
the performances given in Table \ref{tab:split}. In addition, ARTMAP
models have only a single external parameter, and consist of two separate
self-organizing maps, and as such, they do not require any optimization
steps, which renders these family of models to be considerably powerful
in terms of computational time required. The non-iterative nature
of ARTMAP method also enables new observations to be incorporated
to the model as soon as they are obtained, so that the model can be
updated with each new observation without any considerable computational
effort.

The noticeable performance of ARTMAP model compared to traditional
statistical classification techniques, as well as to feedforward multilayer
backpropogation method, particularly in terms of generalizability
over new data sets suggest that ART-based methods, as presented in
this report are potentially robust statistical techniques that can
be used instead of already familiar methods. Considering their relatively
little computational requirements compared to their closest follower
backpropogation models, ART-based models seem to be potential candidates
as future predictive models in ecology.

\bibliographystyle{apalikeNoTitleFormat}
\bibliography{../BIBLIO/tanc}
 \newpage

\section*{Figure Captions}

\begin{description}
\item [Figure~1:]Schematic representation of fuzzy ARTMAP architecture.
Input vectors are processed in ART$_{a}$ module while target categories
are processed in ART$_{b}$ module. Semi-disks represent adaptive
weights. For details, see text. (redrawn from \cite{carpenter:etal1992}). 
\end{description}
\newpage

\section*{Figures}

\newpage

\section*{Tables}

\begin{table}

\caption{Performance of the models on training and test sets. \textit{N}:
number of data points; \textit{P}: number of input variables; k-NN:
k-nearest neighbor; LDA: linear discriminant analysis; QDA: quadratic
discriminant analysis; GLM: generalized linear model; BackProp: feedforward
multilayer backpropogation network; ARTMAP: adaptive resonance theory
based supervised learning. The performance is given as c-index for
backpropogation network, and as percent correctly classified for other
models (see text). }

\begin{center}\begin{tabular}{l|cccccccc}
Set&
N&
P&
k-NN&
LDA&
QDA&
GLM&
BackProp&
ARTMAP\tabularnewline
\hline
\textit{L. senator}(train)&
274&
12&
.828&
.781&
.799&
.859&
1.00&
1.00\tabularnewline
\textit{L. senator}(test)&
273&
12&
.678&
.780&
.798&
.781&
.831&
.971\tabularnewline
\hline
\textit{H. pallida}(train)&
246&
12&
.866&
.488&
.496&
.759&
.874&
1.00\tabularnewline
\textit{H. pallida}(test)&
245&
12&
.669&
.486&
.502&
.703&
.657&
.980\tabularnewline
\hline
\textit{C. brachydactyla}(train)&
294&
12&
.847&
.646&
.701&
.855&
1.00&
1.00\tabularnewline
\textit{C. brachydactyla}(test)&
293&
12&
.765&
.648&
.703&
.769&
.809&
.962\\
\tabularnewline
\end{tabular}\end{center}

\label{tab:split}
\end{table}

\end{document}